\documentclass[aps,reprint,amsmath]{revtex4-2}

\usepackage[utf8]{inputenc}
\usepackage{graphicx}

\begin{document}
\title{Phase-Dependence of Resonant and Antiresonant Two-Photon Excitations}

\author{L. Drescher}\email{lbdrescher@berkeley.edu}\altaffiliation[Current Address: ]{Department of Chemistry, University of California, Berkeley, California 94720, USA}

\author{T. Witting}
\author{O. Kornilov}
\author{M.J.J. Vrakking}

\affiliation{Max-Born-Institut für nichtlineare Optik und Kurzzeitspektroskopie, Max-Born-Strasse 2A, 12489 Berlin, Germany}
\date{\today}

\begin{abstract}
Measurements of the phase of two-photon matrix elements are presented for resonant and antiresonant two-color ionization of helium. A tunable, narrow-bandwidth, near-infrared (NIR) laser source is used for extreme ultra-violet (XUV) high-harmonic generation (HHG). The 15th harmonic of the laser is used within (1+1') XUV+NIR two-photon ionization, and tuned in and out of resonance with members of the 1s$n$p $^1$P$_1$ ($n=3,4,5$) Rydberg series, covering a broad spectral range with high spectral resolution. The technique allows to observe characteristic rapid changes in the phase of the two-photon matrix elements around the resonances and, previously unobserved, at the antiresonances between the resonances. Similar effects are observed for (1+2') XUV+NIR three-photon ionization. The experimental results are compared to a perturbative model and to numerical solutions of the time-dependent Schrödinger equation (TDSE) in the single active electron (SAE) approximation, elucidating the origin and dependences of the observed phenomena.
\end{abstract}

\maketitle

Multi-photon absorption~\cite{goppert-mayer-1931} is a fundamental non-linear response of materials to intense laser radiation, and a key aspect of a wide range of applications of lasers that include four-wave mixing~\cite{dorman-1999,drescher-2021} and HHG~\cite{corkum-1993,schafer-1993}, as well as non-linear microscopy techniques such as Two-Photon-Excited Fluorescence microscopy~\cite{denk-1990}. According to perturbation theory, the efficiency of multi-photon absorption processes is enhanced by the occurrence of intermediate state resonances. On the contrary, between two resonances the excitation amplitudes cancel out, leading to the antiresonant minimum in efficiency. Since the antiresonance condition depends on the precise cancellation of amplitudes, its study allows to measure properties of resonances without exciting them. The antiresonant phase is well explored in e.g. mechanical systems~\cite{wahl-1999} or cavity-coupled quantum systems~\cite{sames-2014,harder-2016}. It provides valuable insight in and characterization of strongly correlated and coupled systems. However, in the domain of multiphoton ionization of atoms and molecules the study of antiresonances have so far been hindered by the difficulties in precise determination of the amplitude minimum~\cite{faisal-1987}. Reconstruction of attosecond beating by interference of two-photon transitions (RABBITT~\cite{muller-2002}) has been successfully used to measure resonant two-photon ionization phases~\cite{swoboda-2010,gruson-2016,kotur-2016,barreau-2019}, but the discussion of the antiresonant phase has so far been absent in this context, due to either limited resolution or insufficient spectral range. In this work, we have extended the well-established broadband RABBITT technique to study antiresonances. We employed narrow bandpass filters to select spectrally narrow laser pulses and tune their central wavelength over the entire broadband spectrum of our laser source. This allowed us to extract RABBITT phases with high spectral resolution and observe antiresonance phase jumps in helium.

The phase of the resonant excitation was elegantly investigated by Dudovich et al.~\cite{dudovich-2001}, who demonstrated that the resonance-enhanced two-photon transition amplitude can be described by a sum of two terms, one describing resonant excitation of an intermediate state in the course of the multi-photon transition, and one describing a near-resonant enhancement mediated by this state. For two-color, XUV+NIR ionization, this gives:
\begin{equation}
\begin{split}
a^\textrm{(near)res}_{g\to f} =& \sum_n \frac{\mu_{fn}\mu_{ng}}{\hbar^2}\Big(-\pi E_\textrm{XUV}(\omega_{ng}) E_\textrm{NIR}(\omega_{fg}-\omega_{ng}) \\ 
&+ i\:\textrm{PV}\int_{-\infty}^{\infty}\textrm d \omega \frac{E_\textrm{XUV}(\omega) E_\textrm{NIR}(\omega_{fg}-\omega)}{\omega_{ng}-\omega}\Big), \label{eq:snd_order}
\end{split}
\end{equation}
where $\omega_{ng}$, $\omega_{fg}$ are transition frequencies connecting the ground and an intermediate state, and the ground and the final state, respectively. $\mu_{ng}$ and $\mu_{fn}$ are the transition dipole moments between these states, and $E_{XUV}(\omega)$ and $E_{NIR}(\omega)$ are the  XUV and NIR spectral amplitudes, respectively. PV denotes the Cauchy Principle Value integral. For the derivation of this expression from time-dependent perturbation theory~\cite{boyd-2013} see SM.
In this expression, the first term describes resonant XUV absorption from the ground to an intermediate state ($g \to n$), utilizing a photon with frequency $\omega_{ng}$, followed by an NIR-induced transition to the final state ($n\to f$), by means of absorption of a photon with a frequency $\omega_{fg}-\omega_{ng}$. The second term, containing the Cauchy Principal Value integral, describes a near-resonant transition utilizing all combinations of XUV and NIR photons whose energy adds up to the energy difference between the ground and final state. An important difference between the two terms is the fact that the first term describing the resonant amplitude is real, whereas the term describing the near-resonant amplitude is imaginary, leading to a difference in phase of the (complex) transition amplitude depending on the frequency-dependent weight of the two terms.
If the XUV frequency is tuned to an intermediate state resonance, the first term (real) in Eq.~\eqref{eq:snd_order} is dominant and the phase of the transition amplitude $\arg\left[a^\textrm{(near)res}_{g\to f}\right]=\pi$.
In contrast, above and below the resonance the second term (imaginary) in Eq.~\eqref{eq:snd_order} is dominant, which is positive when $\omega<\omega_{ng}$ and negative when $\omega>\omega_{ng}$. It follows that the phase of $a^\textrm{(near)res}_{g\to f}$ behaves as:
\begin{subequations}
\begin{eqnarray}
\arg\left[a^\textrm{(near)res}_{g\to f}\right]\to&\frac{\pi}{2}, \qquad&\omega < \omega_{ng}\label{eq:imag_above} \\
\arg\left[a^\textrm{(near)res}_{g\to f}\right]\to&\frac{3\pi}{2}, \qquad&\omega > \omega_{ng}\label{eq:imag_below}.
\end{eqnarray}
\end{subequations}

Approximately midway between two intermediate state resonances, the dominant intermediate state $n$ in the sum in Eq.~\eqref{eq:snd_order} changes and the phase of $a^\textrm{(near)res}_{g\to f}$ promptly changes from being described by Eq.~\eqref{eq:imag_above} to that of Eq.~\eqref{eq:imag_below}. This is the antiresonance.

Multi-photon interferometry plays an essential role in the recently developed field of attosecond science. One of the most popular techniques in attosecond science is RABBITT, which is used to characterize the time structure of attosecond pulse trains (APTs)~\cite{paul-2001,lopez-martens-2005}, as well as photoionization phases of atoms~\cite{klunder-2011,swoboda-2010,gruson-2016,isinger-2017} and molecules~\cite{haessler-2009,caillat-2011,vos-2018,kamalov-2020} that are exposed to such an APT. In RABBITT, XUV ionization in the presence of an NIR field leads to the formation of sidebands (SB) in the photoelectron spectrum (PES), where the total absorbed photon energy is $(q+1)\omega_\textrm{NIR}$ (with $q+1$ being even), described by

\begin{equation}
\begin{split}
I\big((q+1)\omega_\textrm{NIR},\tau\big) &= \Big|a^{(+)}_{g\to f}\big(q\omega_\textrm{NIR}\big)e^{i\omega_\textrm{NIR}\tau} \\
&+a^{(-)}_{g\to f}\big((q+2)\omega_\textrm{NIR}\big)e^{-i\omega_\textrm{NIR}\tau}\Big|^2 \label{eq:rabbitt_def},
\end{split}
\end{equation}
where the two-color, two-photon ionization amplitude $a^{(+)}_{g\to f}$ ($a^{(-)}_{g\to f}$) describes absorption of an XUV harmonic $q$ ($q+2$) accompanied by the absorption (emission) of an NIR photon of frequency $\omega_\textrm{NIR}$.
Coherence of the two pathways $a^{(+)}_{g\to f}$ and $a^{(-)}_{g\to f}$ leads to interference in the final state. The result of the interference (i.e. constructive or destructive) depends on the one-photon dipole phase, the continuum-continuum transition phase, the atto-chirp of the harmonics~\cite{dahlstrom-2012}, and the delay-dependent phase of the NIR field ($\omega_\textrm{NIR}\tau$). The maximum sideband intensity occurs for delays where
\begin{equation}
\arg [a^{(-)}_{g\to f}]-\arg[a^{(+)}_{g\to f}] - 2\omega_\textrm{NIR}\tau = 2\pi \cdot n\label{eq:rabbitt_phase}
\end{equation}

When harmonic $q$ is (near-)resonant with an excited state, $a^{(+)}_{g\to f}(\omega)$ is given by the expression for $a^\textrm{(near)res}_{g\to f}$ in Eq.~\eqref{eq:snd_order}. Consequently, it is expected that the sideband signal allows to determine whether resonant multi-photon absorption as described by the first term in Eq.~\eqref{eq:snd_order} or near-resonant absorption as described by the second term is dominant.

An experimental setup previously described in detail~\cite{neidel-2013,drescher-2018,drescher-2021} was adapted to measure RABBITT signals from (anti)resonant two-photon ionization of helium. To selectively probe resonances at a range of harmonic wavelengths, interferometric bandpass filters were used to extract a narrow spectral region from the broad spectral output of a commercial laser amplifier (1\,kHz repetition rate, 28\,fs FWHM duration pulses, 790\,nm central wavelength). The central wavelength of the selected spectral region was varied by rotating the angle of incidence of the bandpass filter. To achieve a broad range, three different bandpass filters were used (two 10\,nm FWHM bandpass filters, 810\,nm and 790\,nm central wavelength at normal incidence, and one 5\,nm FWHM bandpass filter, 808\,nm central wavelength). In this manner, the central wavelength could be varied from 770\,nm to 806.5\,nm in roughly 1\,nm steps. The spectral narrowing of the pulses led to an elongation of the pulse duration. A wide delay-range RABBITT scan (measured with the 810\,nm filter tuned to 790\,nm) showed a cross-correlation duration of the NIR and XUV pulse of 114\,fs (FWHM).
The beam pointing after the bandpass filter was actively stabilized by a proportional-control loop feeding back deviations of focused and unfocused beam positions, imaged by digital cameras, to two motorized mirror mounts. It was furthermore passively stabilized by focusing the beam into an evacuated hollow-core fiber waveguide. The filtered narrow bandwidth NIR pulses (with pulse energies varying from 1\,mJ to 1.5\,mJ) were separated into two arms of a Mach-Zehnder type interferometer by a beamsplitter mounted on a motorized piezo-stage. One part was used to generate high harmonics by focusing the NIR pulses into an argon cell inside the vacuum system. The generated XUV APT was separated from the driving NIR pulses by a thin aluminium filter (200\,nm thickness). The APT was recombined with a time-delayed replica of the NIR pulse from the second arm of the interferometer, using a holey-mirror. A CW-laser co-propagated through the interferometer and was separately recombined under a small tilt angle. The resulting sheer interferometery fringes were imaged by a digital camera. The delay-dependent fringe pattern was used to generate a feedback signal onto the piezo-motorized stage beneath the beamsplitter to control and stabilize the time-delay in the interferometer. The intensity of the replica NIR pulse was attenuated by a fixed-size iris (2.8\,mm diameter), leading to an NIR intensity in the experimental region of approximately $10^{11}$\,W/cm$^2$, thereby limiting the influence of the AC Stark shift and higher-order NIR processes.
After XUV-NIR recombination, the two-color laser field was re-focused by a grazing-incidence toroidal mirror into a velocity-map imaging spectrometer (VMIS)~\cite{eppink-1997,ghafur-2009}. There, the laser beam intersected a jet of helium atoms from a pulsed gas source. Photoelectrons resulting from two-color XUV$\pm$NIR ionization were imaged onto a microchannel-plate (MCP) + phosphor screen combination and the resulting images were recorded by a digital camera.

\begin{figure}[tb]
\includegraphics[width=.5\textwidth]{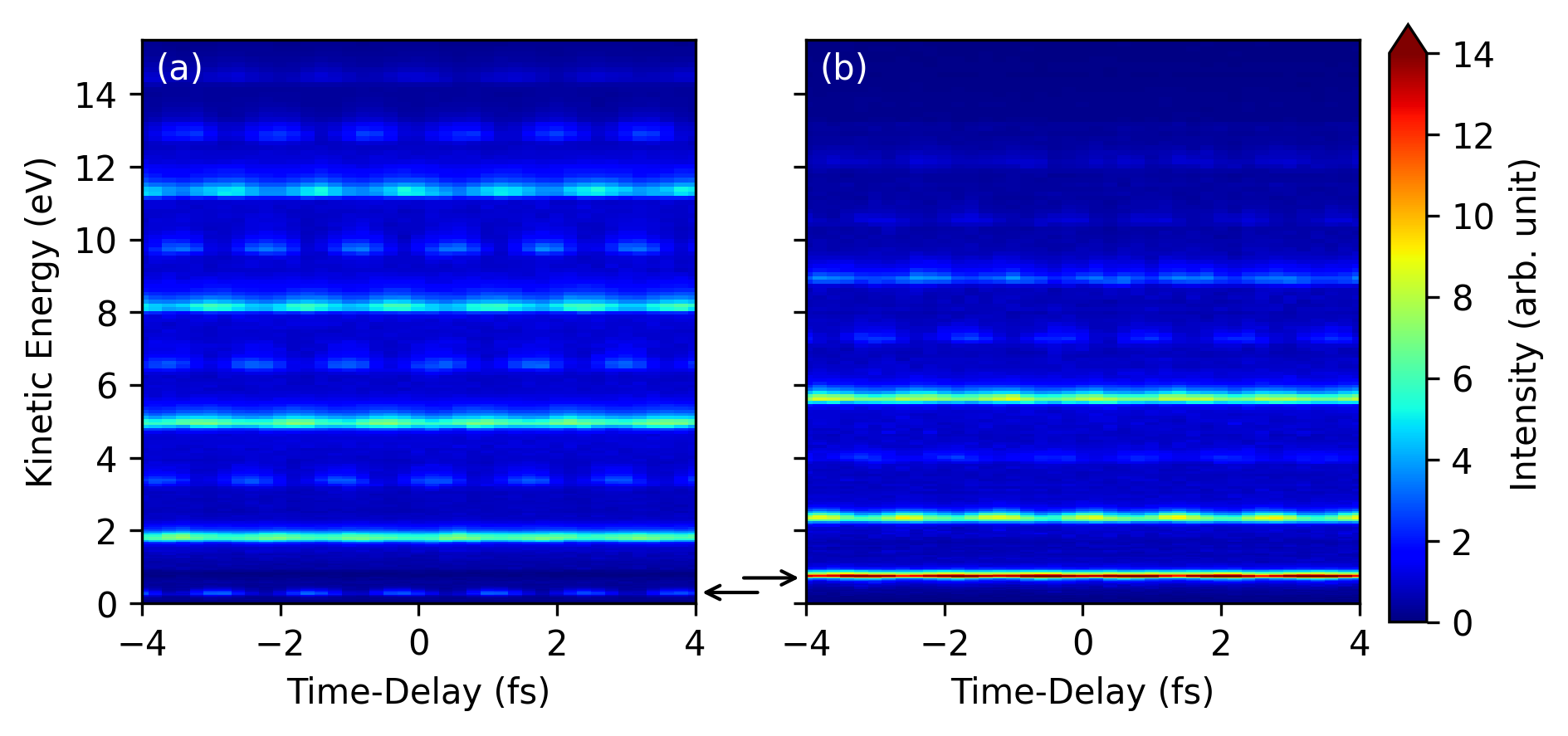}
\caption{Delay-dependent PES for two-color XUV$\pm$NIR ionization of He, using a fundamental wavelength of (a) 795\,nm and (b) 784\,nm. Alternating sideband contributions (SB) and harmonic contributions are observed, the lowest one being that of SB16 (denoted by arrow). When H15 is in (near-)resonance with an excited state of the 1s$n$p series, as is the case in (b), a pronounced resonant-enhancement of the SB16 intensity can be observed.\label{fig:pes}}
\end{figure}

To calculate the photoelectron kinetic energy distribution, an inverse Abel transformation was performed using the BASEX algorithm~\cite{dribinski-2002}. PES were obtained by integrating along the polar-angle between $-\pi/4$ and $0$, as well as between $0$ and $\pi/4$. To determine the phase of the RABBITT signal, the time delay between the XUV and NIR pulses was varied over a range of 20\,fs around zero delay, in 200\,as steps. Typical delay-dependent PES, recorded with an NIR spectrum centered at 795\,nm are shown in Figure~\ref{fig:pes}(a). The contributions from single-photon ionization by the harmonics and from the two-color XUV$\pm$NIR sidebands are readily visible, starting with sideband 16 (SB16) at 0.2\,eV kinetic energy, up to the 25th harmonic (H25) at 15\,eV. As expected in a RABBITT experiment(see Eqs.~(\ref{eq:rabbitt_def},~\ref{eq:rabbitt_phase})), the intensities of the harmonic and SB photoelectron signals oscillate with XUV-NIR time delay, with one full oscillation occurring in half an NIR optical period (1.3\,fs). When H15 is resonant with a Rydberg state, a resonance enhancement of SB16 can be observed, which can then become the dominant contribution to the PES (see Figure~\ref{fig:pes}(b), $\omega_\textrm{NIR} = 784$\,nm). By contrast, in the absence of a nearby resonance, the amplitude of SB16 is comparable to that of SB18 and SB20 (see Figure~\ref{fig:pes}(a)). To avoid damage to the phosphor screen and saturation of the camera, the gain of the MCP detector and the exposure time were adjusted for different filter positions.

A Takeda-type algorithm~\cite{takeda-1982} was employed to calculate the phase of the sideband and harmonic oscillations: the delay-dependent PES were spectrally filtered in Fourier domain by using a Gaussian filter centered at twice the (positive) fundamental frequency ($2\omega_\textrm{NIR}$) and RABBITT phases (i.e. the phases of the oscillations of chosen sidebands and harmonic peaks) were then calculated from the complex-valued back-transformation of the filtered Fourier spectrum (at an arbitrary point of the delay-axis).

Since variations in the intensity of the NIR driving field can cause shifts in the photon energy of the harmonics~\cite{wahlstrom-1993}, an effective driver laser frequency was calculated by evaluating the first moment of the peak in the PES associated with SB16, H17, SB18 and H19.
For each filter position the delay scan was repeated four times. Moreover, the entire measurement was performed two times for the 810\,nm/10\,nm filter and three times for the 790\,nm/10\,nm and 808\,nm/5\,nm filter. Due to a slight alignment deviation of the VMIS, only the upper half of the photoelectron momentum distribution was evaluated, for which the best spectral resolution could be achieved. By independently evaluating the upper left and upper right quadrant, 24 distinct RABBITT phase determinations were obtained (16 in the case of the 810\,nm/10\,nm filter), which permit an estimation of the confidence intervals (CI) of the effective driver laser frequency and RABBITT phase.

\begin{figure}[tb]
\includegraphics[width=.5\textwidth]{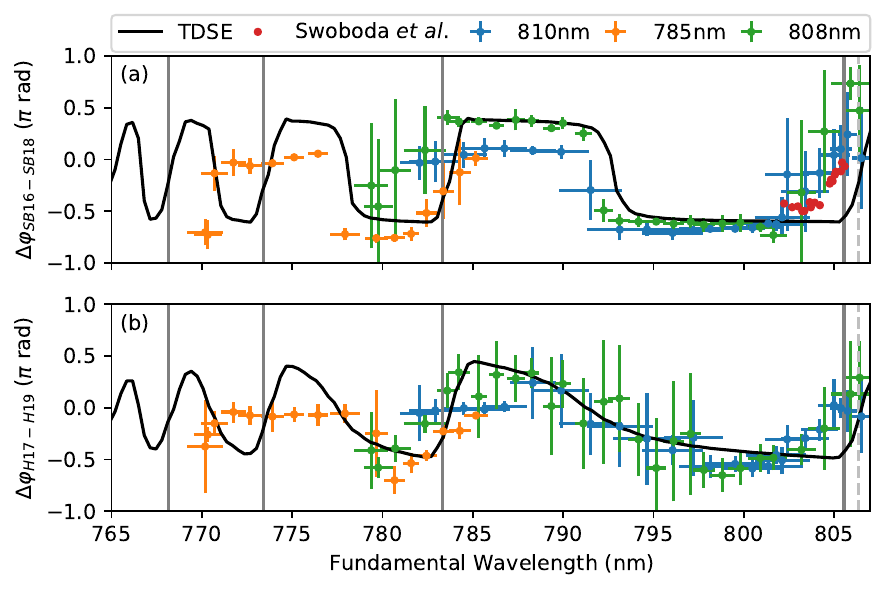}
\caption{Measured and calculated wavelength dependence of the phase difference between RABBITT oscillations at the energies corresponding to (a) SB16 and SB18 and (b) H17 and H19, respectively, revealing the influence of resonances in (1+1') and (1+2') XUV+NIR ionization. The points correspond to experiments performed with different bandpass filters: $\lambda =$ 810\,nm (blue), 808\,nm (green) and 790\,nm (orange). TDSE simulations are shown as a solid black curve. Vertical grey bars indicate fundamental wavelengths for which H15 is resonant with members of the 1s$n$p $^1$P$_1$ series of He~\cite{nist}. The SAE-potential leads to a lower resonant fundamental wavelength for the TDSE (dashed light grey) The errorbars indicate the standard deviation over several measurements (see main text). For comparison, the previously published results of Swoboda \textit{et al.}~\cite{swoboda-2010} are shown as red dots in (a).\label{fig:sb16}}
\end{figure}

In Figure~\ref{fig:sb16}(a) the measured difference between the RABBITT phase of SB16 and SB18 is shown as a function of the effective driver laser wavelength. This phase difference was evaluated, since the amplitude of the H15+NIR ionization cannot be measured without its interference with the H17-NIR amplitude, and moreover since the precise zero time-delay between XUV and NIR pulses is not available from the experiment. A change in phase of about $\pi$ is observed around wavelengths for which H15 is resonant with one of the 1s$n$p $^1$P$_1$ ($n=3,4,5$) states (grey vertical lines in Figure~\ref{fig:sb16}(a)). The results are consistent with phase measurements for wavelengths above 799\,nm, which were previously obtained by Swoboda \textit{et al.}~\cite{swoboda-2010} (red dots in Figure~\ref{fig:sb16}(a)). In addition to the phase changes at the resonance positions, a second set of pronounced phase changes can be observed approximately midway between two resonances, indicative of antiresonant behavior. The wavelength dependence of these changes are consistent with the simple perturbative model described in Eq.~\eqref{eq:snd_order} and Eq.~\eqref{eq:rabbitt_def}. Note that (by convention) the (near-)resonant two-photon excitation phase enters the RABBITT phase in the experiment with its sign inverted (see Eq.~\eqref{eq:rabbitt_phase}). An increase in the CI is observed at wavelengths where the signal contrast is diminished, either due to a geometrical limit of the bandpass filter angle or due to resonant enhancement of (1+1') XUV+NIR photoionization, which required a reduced gain and led to an increase in background signal. In contrast, close to antiresonance, the RABBITT signal can be measured with high contrast and therefore narrow CI. 

To further corroborate these results and to connect measured relative phases to predictions of the perturbative model, TDSE simulations were performed using code described in Ref.~\cite{muller-1999}. The TDSE was solved in single-active electron approximation (SAE) on a 2D grid (radial distance and angular momentum). Detailed parameters of the calculation can be found in the SM. An underestimation of the 1s3p resonant energy in the SAE potential led to a deviation of the observed resonant phase in Figure~\ref{fig:sb16}.
\begin{figure}[tb]
\includegraphics[width=.5\textwidth]{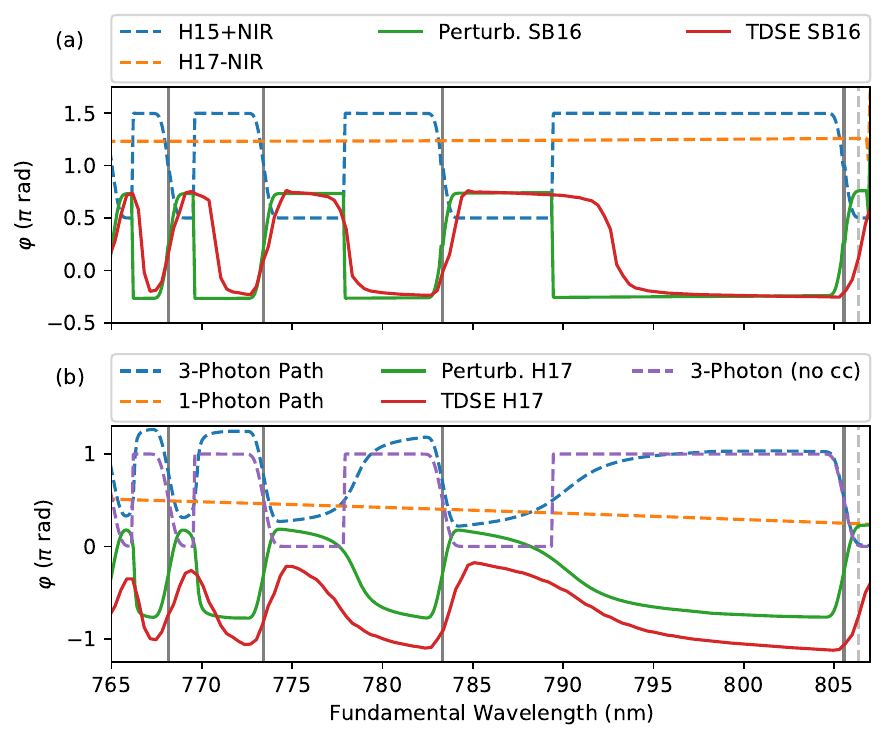}
\caption{Comparison of the SB16 phase (a) and the H17 phase (b) calculated by solving the TDSE (red) and by the perturbative model (green). (a) Two pathways contribute to the SB16 signal in the perturbative model, namely the H15+NIR (dashed blue) and the H17+NIR (dashed orange) pathway. (b) Similarly, the phase of the H17 signal is determined by two pathways, namely a three-photon pathway including H15 (dashed blue) and the direct one-photon ionization by H17 (dashed orange). When the continuum-continuum transition phase is omitted
from the three-photon pathway (dashed purple), the smooth wavelength-dependence of the H17 phase between two resonances is no longer observed.\label{fig:tdse_perturbation}}
\end{figure}

The phases of the numerical PES were evaluated with the same method that was applied to the experimental results (see above). For the analysis, the PES were integrated over all angles. 
The wavelength dependence of the phase difference between SB18 and SB16 from the TDSE calculation is shown in Figure~\ref{fig:sb16}(a). A good agreement with the experimental results is observed above a wavelength of 780\,nm. 
Since the zero XUV-NIR time-delay is known for the TDSE results, absolute RABBITT phases of SB16 and SB18 can be extracted, and are shown in Figure~\ref{fig:tdse_perturbation}(a), supporting that the observed wavelength-dependence in Figure~\ref{fig:sb16}(a) originates exclusively from the SB16 phase. Figure~\ref{fig:tdse_perturbation}(a) also contains predictions for the SB16 phase from  the perturbative model in Eq.~\eqref{eq:snd_order}. Transition wavelengths were taken from literature values~\cite{nist} and dipole moments were calculated using several analytical approximations~\cite{yatsenko-1999,matsumoto-1991,veniard-1990}. Calculations were limited to the dominating angular-momentum channels~\cite{chang-1995,busto-2019}. Details of the model calculation can be found in the SM.

The resulting SB16 phase, as well as phases for the H15+NIR and the H17-NIR pathway are compared in Figure~\ref{fig:tdse_perturbation}(a). They show good agreement with results from TDSE calculations. The largest deviations are observed at the antiresonance between two resonances. These deviations can be rationalized in terms of different strengths of the transition dipole moments as used in the perturbative model, to which the antiresonant condition is highly sensitive~\cite{faisal-1987}. Note that the H17-NIR pathway includes an NIR-driven continuum-continuum (cc) transition. RABBITT measurements are typically conducted to determine the phase difference between two adjacent harmonics which each are energetic enough to ionize, leading to two cc contributions~\cite{dahlstrom-2013,harth-2019}. In the experiment described here, however, the cc transition only contributes to the H17-NIR pathway, thus providing direct experimental access to the magnitude of this phase. The model predicts that the one-photon dipole and cc transition phase combined, i.e. the phase of the H17-NIR pathway, add a phase of $\approx\frac{5}{4}\pi$ to the SB16 RABBITT phase (see Figure~\ref{fig:tdse_perturbation}(a)), while the contribution of the cc phase alone increases from $\approx\frac{3}{4}\pi$ to $\approx\pi$ (between $\lambda_\textrm{NIR}=765$\,nm and 806\,nm, not shown).

A wavelength-dependent phase can additionally be observed in the signal corresponding to H17 absorption. Photoelectrons directly emitted by H17 interfere with those from H15+2NIR three-photon ionization, leading to a delay-dependent RABBITT signal. In Figure~\ref{fig:sb16}(b) the phase difference between the oscillations in the H17 and the H19 signal is shown for experimental results and TDSE calculations. Again, a good agreement is observed for wavelengths larger than 780\,nm. The data once more reveal the clear influence of (near)-resonant enhancement of the pathway involving H15. Compared to the SB16 signal, a smoother transition of the RABBITT phase is observed between two resonances. The perturbative model is further developed for the H17 phase (see SM) and the result shown in Figure~\ref{fig:tdse_perturbation}(b). While the perturbative model qualitatively reproduces the wavelength dependence of the H17 phase, it shows a slight offset of $\approx0.2\pi$ compared to the TDSE calculation. 
The perturbative model shows that the smoother wavelength-dependence is due to the cc transition phase. The complex cc transition dipole moment leads to a small additional real part to the otherwise purely imaginary near-resonant transition amplitude, as was previously the case in the H15+NIR pathway.
As the wavelength crosses the antiresonance, the amplitude never becomes exactly zero owing to this small real part. Consequently the phase does not experience a jump, but smoothly transitions from the $\frac{\pi}{2}$ to $\frac{3\pi}{2}$ values that are seen close to the resonances.
Indeed, omission of the continuum-continuum transition phase in the perturbative modeling for the H17 phase leads to the same abrupt change of phase that was previously observed for SB16 (see Figure~\ref{fig:tdse_perturbation}(b)).

The phase jumps at antiresonances observed in this work are reminiscent of the antiresonances studied for single-atom cavity QED in Ref.~\cite{sames-2014}. In this work the authors tuned the excitation laser wavelength across the antiresonance between the atomic resonance and the cavity resonance. They observed the phase behavior similar to that shown in Fig.~\ref{fig:sb16}. They were able to modify the system parameters affecting the phase shape at the antiresonance and thus could study the properties of the strongly-coupled (atom-cavity) system without exciting it, i.e. non-destructively. We propose, that the method and observations presented here can be applied to naturally strongly coupled resonances in atomic and molecular systems opening a new non-destructive method to study their non-adiabatic and light-induced coupling of states.
In conclusion, we have developed a highly tunable RABBITT-scheme, which allows to observe the phase of wavelength-dependent two-photon ionization of a multi-level system. The variation of the fundamental wavelength allows to characterize RABBITT phases over a broader range of photon energies as in previous experiments~\cite{swoboda-2010,kotur-2016,barreau-2019}. The approach may be seen as complementary to the recent development of RAINBOW RABBITT~\cite{gruson-2016,busto-2018,fuchs-2021}. The combination of both approaches has the potential to increase the spectral resolution or to reduce requirements on the spectrometer resolution.
Two regions could be identified where the RABBITT phase rapidly changes as a function of wavelength, namely one region where a change occurs in the dominance of the resonant, respectively near-resonant pathway, and a second region in between two resonances, where the amplitudes from two neighboring near-resonant pathways cancel out and leads to the antiresonance. One advantage of measuring the antiresonant phase is clear from our results: The absence of a resonant-enhanced background from (1+1') XUV+NIR photoionization leads to a reduced uncertainty on the measured phase. 
By choosing high-harmonic energies which are close to the ionization threshold, we could furthermore study the role of the continuum-continuum phase in RABBITT. Similar photoionization phases close to the ionization threshold have recently been measured by Sansone \textit{et al.}~\footnote{G. Sansone, private communication}. Further development of our method would allow to measure the RABBITT phase across the ionization threshold, as recently proposed by Kheifets and Bray~\cite{kheifets-2021}.

\begin{acknowledgments}
The authors would like to thank A. A. Ünal for his support with the laser system and N. Mayer for helpful discussion.
\end{acknowledgments}

\onecolumngrid
\section{Supplementary Material}
\subsection{Parameters of TDSE calculations}
For solving the TDSE, the grid was composed of two dimensions, with one grid dimension corresponding to the distance between the (photo)electron and the He$^+$ core (r = 0..9000\,a.u. in 0.15\,a.u. steps), and one to the (photo)electron orbital angular momentum (l = 0..9). The central wavelength of the 20-cycle FWHM long sin-squared NIR pulse was varied between 0.0607\,a.u. (750\,nm) and 0.0555\,a.u. (820\,nm), in steps of $2.5\cdot 10^{-5}$\,a.u. The XUV pulse consisted of a superposition of 5 harmonics (H11-H19) with a Gaussian temporal profile with a FWHM equal to that of the NIR pulse. The peak electric field of the NIR was 0.0004\,a.u., corresponding to an intensity of $5.6\cdot10^{9}$\,W/cm$^2$ while the peak field strength of each of the 5 harmonics was\,0.001 a.u. The delay between the XUV and NIR pulses was varied between -80 and 80\,a.u. in 4\,a.u. steps. PES were calculated for kinetic energies between 0 and 0.2\,a.u., with a resolution of $2.5\cdot10^{-4}$\,a.u.

\subsection{Parameters of perturbative model calculation}
For the numerical solution of the perturbative model, transition wavelengths and dipole moments for the 1s$n$p $^1$P$_1$ ($n=3...7$) series were taken from the NIST atomic spectra database~\cite{nist}. Transition dipole moments for bound-to-continuum transitions were calculated using a quantum-defect model~\cite{yatsenko-1999}.
For the H17-NIR pathway, transition dipole moments were calculated using hydrogenic models for ionization from the ground state~\cite{matsumoto-1991} and for continuum-continuum transitions~\cite{veniard-1990}. Since the phase of this pathway is expected to only weakly depend on the wavelength~\cite{swoboda-2010}, only one intermediate state at the central wavelength of H17 was included in the calculation. Furthermore, for both pathways the perturbative model was limited to the $D$ final angular momentum channel, which is expected to be dominant due to oscillator strength~\cite{chang-1995} and Fano propensity rules~\cite{busto-2019}. In accordance with experiment, Gaussian NIR pulses were used with a spectral width of 10\,nm (FWHM). The width of the individual harmonics was estimated from the measured XUV spectra as 0.65\,nm FWHM (140meV). 

\subsection{Derrivation of two- and three-photon amplitudes from time-dependent perturbation theory}
To derrive the expression for the two- and three-photon excited final state amplitude, we employ time-dependent perturbation theory.
The amplitude of an $N$-th order multiphoton excitation is described recursevly as:~\cite{boyd-2013}
\begin{equation}
a_m^{(N)}(t) = (i\hbar)^{-1} \sum_l \int_{-\infty}^t dt' V_{ml}(t') a_l^{(N-1)}(t' ) e^{i\omega_{ml}t'},
\end{equation}
where $a_l^{(N-1)}$ is the excitation amplitude of intermediate states $l$ in $(N-1)$-st photon order, $\omega_{ml}=\omega_m-\omega_l$ is the transition frequency between states $l$ and $m$. $V_{ml}(t)=\mu_{ml}E(t)$ is the time-dependent perturbation of the system by the electric field $E(t)=\int_{-\infty}^{\infty}E(\omega) e^{i\omega t}dt$ and $\mu_{ml}$ is the transition dipole moment between states $l$ and $m$. It is assumed that the continuum is sufficiently flat, such that a description of discrete states can be kept throughout bound and continuum states.

Starting with a system in the ground state $g$, i.e. $a_l^{(0)}  = \delta_{lg}$, the two-photon transition amplitude to a state $f$ via intermediate state $n$ is:
\begin{equation}
a_{f}^{(2)}(t) = -\sum_n \frac{\mu_{fn}\mu_{ng}}{\hbar^2} \int_{-\infty}^t dt_1 \int_{-\infty}^{t_1}dt_2 E_\textrm{XUV}(t_2) E_\textrm{NIR}(t_1) e^{i\omega_{ng} t_2} e^{i\omega_{fn} t_1},\label{eq:snd_order_time}
\end{equation}
where we assume that $\omega_{ng}\gg\omega_{fn}$, such that we can safely factorize our two-color field in its components $E_\textrm{XUV}(t)$ and $E_\textrm{NIR}(t)$. 
To simplify the discussion it is assumed that the electric fields are transform limited, i.e. they can be described as real and positive. We also omit the explicit mention of the (trivial) time-delay $\tau$ phase of the NIR field with respect to the XUV field, i.e.:
\begin{equation}
E_\textrm{NIR}(\omega,\tau) = E_\textrm{NIR}(\omega) e^{i\omega\tau}.
\end{equation}

Using the spectral representation of the two electric fields, we can rewrite Eq.~\eqref{eq:snd_order_time} as:
\begin{equation}
a_{f}^{(2)}(t) = -\sum_n \frac{\mu_{fn}\mu_{ng}}{\hbar^2} \int_{-\infty}^{\infty}d\omega_1 \int_{-\infty}^\infty d\omega_2 E_\textrm{XUV}(\omega_2) E_\textrm{NIR}(\omega_1) \int_{-\infty}^t dt_1 \int_{-\infty}^{t_1}dt_2  e^{i(\omega_{ng}-\omega_2) t_2} e^{i(\omega_{fn} - \omega_1) t_1},
\end{equation}
and directly integrate up to $t_1$:
\begin{equation}
a_{f}^{(2)}(t) = -\sum_n \frac{\mu_{fn}\mu_{ng}}{\hbar^2} \int_{-\infty}^{\infty}d\omega_1 \int_{-\infty}^\infty d\omega_2 E_\textrm{XUV}(\omega_2) E_\textrm{NIR}(\omega_1) \int_{-\infty}^t dt_1   e^{i(\omega_{fn}-\omega_1) t_1} \frac{e^{i(\omega_{ng} - \omega_2) t_1}}{i(\omega_{ng} - \omega_2)}.\label{eq:snd_order_spec}
\end{equation}

We will assume that we measure the photoelectron long after the interaction with the electric fields is over and therefore take $t\to\infty$. Using the Fourier property of the Dirac-delta function: $\delta(\omega-\omega_l) = \int_{-\infty}^\infty dt \exp{[i(\omega+\omega_l)t]}$, we can solve the remaining time-dependent integral in Eq.~\eqref{eq:snd_order_spec} and using the integral properties of the Dirac-delta the spectral integral of $\omega_1$, leaving us with:
\begin{equation}
a_{f}^{(2)} = -\sum_n \frac{\mu_{fn}\mu_{ng}}{i\hbar^2} \int_{-\infty}^{\infty}d\omega_2  E_\textrm{XUV}(\omega_2) E_\textrm{NIR}(\omega_{fg}-\omega_2)    \frac{1}{(\omega_{ng} - \omega_2)}	
\end{equation}
Introducing a (negative) damping term to account for limited lifetime $\omega_{ng}=\omega_{n}-\omega_g-i\Gamma$ allows applying the Sokhotski–Plemelj theorem for the resonant case $\omega_2 = \omega_{ng}$ and yields Eq.~(1) in the main text.

We apply similar steps to solve the three-photon transition amplitude to a state $f$ via intermediate states $n$ and $m$.
\begin{equation}
\begin{split}
a_{f}^{(3)} =& \sum_{n,m}\frac{\mu_{ng}\mu_{mn}\mu_{fm}}{i\hbar^3}  \int d\omega_3 \frac{E_{XUV}(\omega_3)}{(\omega_{ng} -\omega_3)}\\
 &\left(i\pi E_{NIR}(\omega_{mg}-\omega_3) E_{NIR}(\omega_{fm}) + \textrm{PV}\int d\omega_2 E_{NIR}(\omega_2) E_{NIR}(\omega_{fg}-\omega_2-\omega_3) \frac{1}{(\omega_{mg} -\omega_2 - \omega_3)}\right).
\end{split}
\end{equation}
As described in the main text, in the continuum we will only consider one resonant excited state $m$ with a resonant frequency $\omega_{mn}$ at the center of the NIR spectrum. Since the principle value integral is then always symmetric around the resonance it evaluates to zero. We can then apply again the Sokhotski–Plemelj theorem to the remaining integral of $\omega_3$:
\begin{equation}
\begin{split}
a_{f}^{(3)}\Big\rvert_m \approx& \sum_{n}\frac{\mu_{ng}\mu_{mn}\mu_{fm}}{i\hbar^3}  \\
&\left[-\pi^2 E_{XUV}(\omega_{ng}) E_{NIR}(\omega_{mn}) E_{NIR}(\omega_{fm}) + i\pi \textrm{PV} \int d\omega_3 \frac{E_{XUV}(\omega_3)}{(\omega_{ng} -\omega_3)}E_{NIR}(\omega_{mg}-\omega_3) E_{NIR}(\omega_{fm}) \right].
\end{split}
\end{equation}

The numerical results of the three-photon excitation are shown in Fig 3(b) in the main text. The parameters are described in the main text. Again two pathways contribute to the H17 signal in the perturbative model, the three-photon pathway containing (near-)resonant excitation of the 1s$n$p $^1$P$_1$ series in He photoionization by an NIR photon and a continuum-continuum transition driven by absorption of an NIR photon and the one-photon pathway of direct photoionization by the 17th harmonic. As described in the main text, the wavelength dependence of the TDSE H17 signal is well reproduced, but the perturbative model shows a varying offset from the TDSE results, indicating the limits of our simple modeling.

\twocolumngrid
\bibliography{main_arxiv.bib}

\end{document}